\documentclass[aps,pre,amsmath,amssymb,lengthcheck,showpacs,superscriptaddress]{revtex4-1}
\usepackage{amsmath,amssymb,bm}
\usepackage[]{graphicx}
\usepackage[]{graphics}
\usepackage{epsfig}
\begin{document}
\title{
Inhomogeneous Transport in Model Hydrated Polymer Electrolyte Supported Ultra-Thin Films  
}
\author{Daiane Damasceno Borges}
\author{Alejandro A. Franco}
\altaffiliation{Current address: Laboratoire de R\'eactivit\'e et de Chimie des Solides (LRCS) - UMR 7314 (CNRS, Universit\'e de Picardie Jules Verne), 80039 Amiens, France}
\affiliation{CEA, DRT/LITEN/LCPEM, 17 rue des Martyrs, 38054 Grenoble Cedex 9, France }
\author{Kourosh Malek}
\affiliation{National Research Council of Canada, Energy, Mining and Environment, Vancouver, BC, Canada}
\author{Gerard Gebel}
\author{Stefano Mossa}
\email{stefano.mossa@cea.fr}
\affiliation{INAC/SPrAM (UMR 5819 UJF, CNRS, CEA), CEA-Grenoble, 17 Rue des Martyrs, 38054 Grenoble, France}
\begin{abstract}
Structure of polymer electrolytes membranes, {\em e.g.}, Nafion, inside fuel cell catalyst layers has significant impact on the electrochemical activity and transport phenomena that determine cell performance. In those regions, Nafion can be found as an ultra-thin film, coating the catalyst and the catalyst support surfaces. The impact of the hydrophilic/hydrophobic character of these surfaces on the structural formation of the films and, in turn, on transport properties, has not been sufficiently explored yet. Here, we report about classical Molecular Dynamics simulations of hydrated Nafion thin-films in contact with unstructured supports, characterized by their global wetting properties only. We have investigated structure and transport in different regions of the film and found evidences of strongly heterogeneous behavior. We speculate about the implications of our work on experimental and technological activity.
\end{abstract}
\maketitle
\section{Introduction}
Polymer electrolyte fuel cells~\cite{kreuer1997,kreuer2001,kreuer2004transport,eikerling2008proton,dupuis2011} are complex devices constituted by soft and hard materials, which are characterized by heterogeneous nano-structure and transport properties. In particular, the membrane-electrode assembly has a strong impact on the fuel cell performance. It is formed, together with other components of limited interest here, by the polymer electrolyte membrane and the catalyst layer. The membrane must be an insulator against electrons transport, promote proton conductivity also in low hydration and high temperature conditions, and be mechanically stable. The most successful material used in these days is Nafion~\cite{mauritz2004state} which, upon hydration, phase-separates in hydrophobic and hydrophilic domains. The former provide mechanical strength, while the latter result in regions available for transport of water and protons. While the macroscopic structure of Nafion has been probed in depth, some details of the nano-structure still remain controversial~\cite{gierke1981morphology,rubatat2004,schmidt2008parallel}. 

The catalyst layer~\cite{zhang2008pem} is the interface region, whose complexity arises from the fabrication process. Here, carbon supported platinum nanoparticles mix with the ionomer and self-organize into extremely heterogeneous structures. Despite significant efforts in the understanding of the bulk Nafion morphology, only a few studies have been devoted to the structure of the hydrated ionomer ultra-thin films formed at the interfaces with platinum and carbon. Indeed, the ionomer is expected to self-organize in different forms, depending on substrate properties such as chemical composition, geometry and, ultimately, wetting behavior. The features of the film have significant impact on many processes including, among others, the electrochemical double layer structure~\cite{franco2006,franco2007a} with the associated charge distribution, 
proton conductivity~\cite{malek2011}, diffusion of reactants and products, degradation kinetics of platinum and carbon support which, in turn, can impact the effective electro-catalytic properties~\cite{francobook2013}.

Computer simulations have been used for a better understanding of the catalyst layer structure. Studies relevant in the present context include Coarse-Grained Molecular Dynamics simulations of the self-organized structure of the catalyst layer~\cite{malek2007self,malek2011a}; atomistic modeling of a) ionomer and water adsorption on graphitized carbon sheets~\cite{mashio2010} and b) interfaces where the ionomer is in contact with the vapor phase and with the catalyst or the catalyst-support surfaces~\cite{liu2008}; simulation of the effect of hydration of Nafion in the presence of platinum nanoparticles~\cite{selvan2012}. 
All these works are based on very detailed modeling of the solid phase in contact with the hydrated ionomer, both at the level of nano-structure and heterogeneity of interactions. 

Here we have chosen a different approach, which focuses very little on chemical details but aims at grasping the main general physical features. This strategy has been successfully applied in studies of molecular liquids, like pure water in contact with infinite structured walls~\cite{giovambattista2007}, or even simple atomic supercooled liquids~\cite{scheidler2002,scheidler2004}. We have considered a mean-field-like interaction ionomer/substrate, that allows us to precisely control the hydrophilic character of the substrate by using a unique tunable control parameter. By using classical Molecular Dynamics simulations, we have explored self-assembly and thin-film formation of a detailed model for the Nafion ionomer, when placed in contact with unstructured infinite hydrophobic and hydrophilic surfaces (supports), under high hydration conditions. Also, we have focused on transport properties of water molecules, and found evidences of strongly heterogeneous behaviour in different regions of the film. Here we report our main findings, further details~\cite{proceedings} and more extended data sets will be included in other publications (Damasceno Borges, D. unpublished).

In what follows, we describe the interaction model between the hydrated ionomer and the support. Next, we report and discuss our main results, including film conformation and transport properties of water molecules in different regions of the film. We conclude with a summary of our findings and perspectives of our work. Details of the computer simulation procedures and some calculations are given in the Methods section.
\begin{figure}[t]
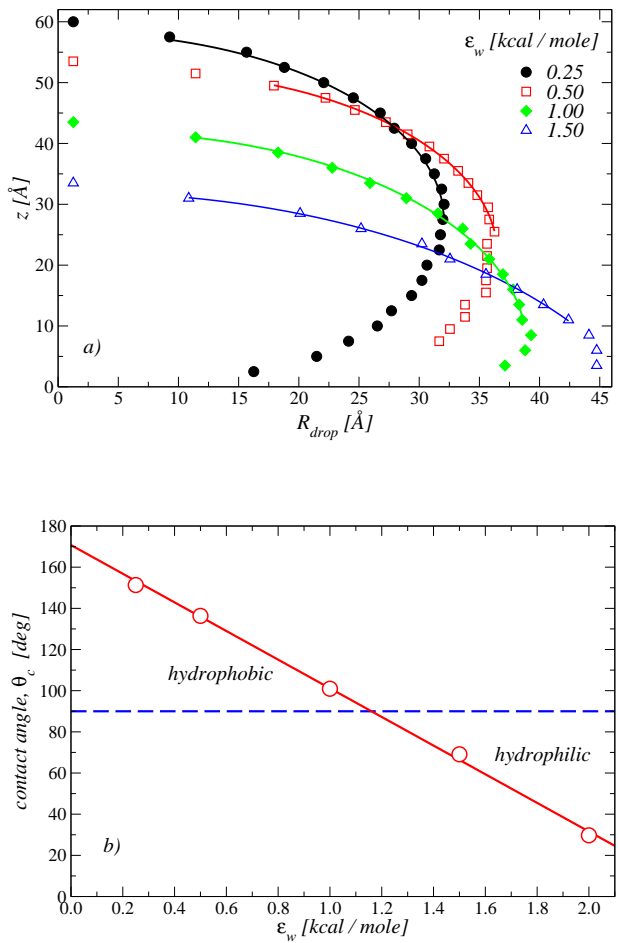

\centering
\includegraphics[width=0.45\textwidth]{drop.eps}
\vspace{1.0cm}

\includegraphics[width=0.45\textwidth]{contact.eps}
\caption{
a) Contour shape of droplets of water molecules deposited on supports characterized by the indicated values of $\epsilon_w$. These data can be directly fit (solid lines) to determine the contact angles $\theta_c$, which are shown in b). Details of these calculations are given in the Methods section.} 
\label{fig:contact}
\end{figure}

\section{Results and Discussion}
{\bf Model.} The Nafion ionomer~\cite{mauritz2004state} is composed by a hydrophobic poly-tetrafluoroethylene backbone with intercalated perfluorinated side chains, which are terminated by a hydrophilic $SO_3^-$ group. An united-atom description is used for representing $CF_{2,3}$ groups in the polymer chain. In contrast, radical sulfonic acids, water molecules and hydronium ions are described with all-atoms resolution. This hybrid modeling scheme is largely used to represent complex polymer systems. Intra-molecular bonded interactions include bonds, angles, and dihedrals terms. Non-bonded interactions are described in terms of Lennard-Jones potentials and long-range Coulombic interactions. Potential parameters are similar to those of Refs.~\cite{venkatnathan2007,masuda2009}. Water molecules are described by the rigid SPC/E model~\cite{berendsen1987}, while the flexible model for the hydronium ion has been taken from Ref.~\cite{kusaka1998}.

All system units interact with a smooth unstructured wall (the support), placed at $z=0$ and parallel to the $xy$-plane, {\em via} a $9\text{--}3$ Lennard Jones potential. This only depends on the distance, $z$, of the unit from the support:
\begin{equation}
V_{w}^\alpha(z)=\epsilon^\alpha_w\left[ \frac{2}{15}\left(\frac{\sigma_w^\alpha}{z}\right)^9-\left(\frac{\sigma_w^\alpha}{z}\right)^3\right]\theta(z_c-z),
\label{equation:wall}
\end{equation}
where $z_c=$~15\AA\ is a cut-off distance and $\theta$ is the Heaviside function. The index $\alpha$ identifies complexes ($H_2O$, $H_30^+$, $SO_3^-$) with significant dipolar coupling to the (hydrophilic) support ($\alpha=\text{phyl}$), or units corresponding to the hydrophobic sections of the polymer ($\alpha=\text{phob}$) which, in contrast, interact very mildly.  The energy well $\epsilon_w^\text{phob}=0.25$~kcal/mole, while $\epsilon_w^\text{phyl}=\epsilon_w$ is our control parameter. The typical interaction length scale $\sigma_w^\alpha=3.2$\AA\ in all cases. 

We have calculated for $\epsilon_w\in[0.25,2]$~kcal/mole the values of the contact angle, $\theta_c$, of a droplet of water molecules interacting with the support {\em via} Eq.~(\ref{equation:wall}), and demonstrated that this range encompasses hydrophobic to strongly hydrophilic behaviour (lower to higher $\epsilon_w$, respectively). We have used the technique described in~\cite{shi2009molecular,werder2003water}, based on a  direct fit of the contour shape of the droplet. These data are shown as symbols in Fig.~\ref{fig:contact}a). A circular best fit through these points (solid lines) is extrapolated to the wall surface where the contact angle, $\theta_c$, is measured (Fig.~\ref{fig:contact}b)). Details of these calculations are given in the Methods section.
\begin{figure}[t]
\centering
\includegraphics[width=0.49\textwidth]{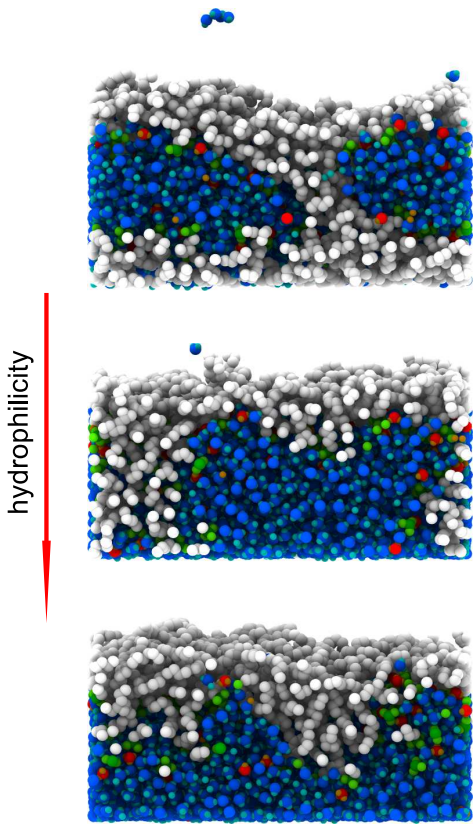}
\caption{
Snapshots of hydrated Nafion ultra-thin films, at $T=350$~K and $\lambda=22$,  for an interaction with the support of increasing hydrophilic character ($\epsilon_w=0.25, 1.0, 2.0 $~kcal/mole, from top to bottom). We observe the formation of extended water pools (blue) which are separated from the confining polymer matrix (grey) by the charged sulfonic groups interface (green); hydronium complexes are also shown (red). For $\epsilon_w=2.0$~kcal/mole the ionomer is completely desorbed from the substrate. Note the evaporated water molecules, on the top of the films.}
\label{fig:snapshots}
\end{figure}

\begin{figure}[b]
\centering
\includegraphics[width=0.45\textwidth]{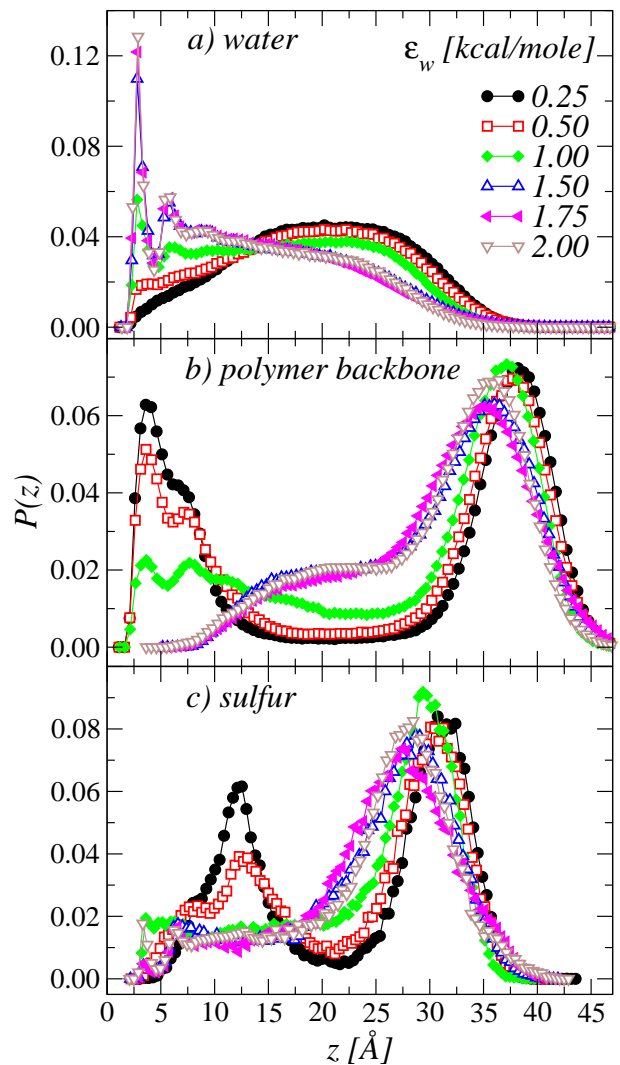}
\caption{
Mass probability distributions as a function of the distance from the support, $z$, at the indicated values for $\epsilon_w$. We have considered a) water oxygens, b) polymer backbone units, and c) sulfur atoms. 
}
\label{fig:masses}
\end{figure}

{\bf Thin film morphology.} In Fig.~\ref{fig:snapshots} we show typical snapshots of the self-organized ionomer ultra-thin film, for interactions with the support of increasing hydrophilic character (from top to bottom). We observe the hydrophobic character of the ionomer/air interface~\cite{bass2010} and the formation of extended hydrophilic pools, separated from the confining polymer matrix by an interface decorated with the sulfonate groups. Significant re-organization of the ionomer on increasing $\epsilon_w$ is also visible. In order to quantify these changes, we plot in Fig.~\ref{fig:masses} the normalized mass probability distributions for a) water, b) polymer backbone and c) sulfur atoms. For the most hydrophobic cases we have found a sandwich structure, with a wide water distribution centered at $z\simeq 20$\AA\ and most part of the polymer in contact with the support and at the open interface with air. Distribution of the sulfonate groups coherently shows two peaks at the ionomer/water interfaces. By increasing $\epsilon_w$, the ionomer abruptly desorbs from the support, inducing a significant re-organization of water domains. Water completely floods the region close to the support, where density layering is observed. A substantial decrease of hydration in regions far from the support is an expected consequence.  

\begin{figure}[t]
\centering
\includegraphics[width=0.49\textwidth]{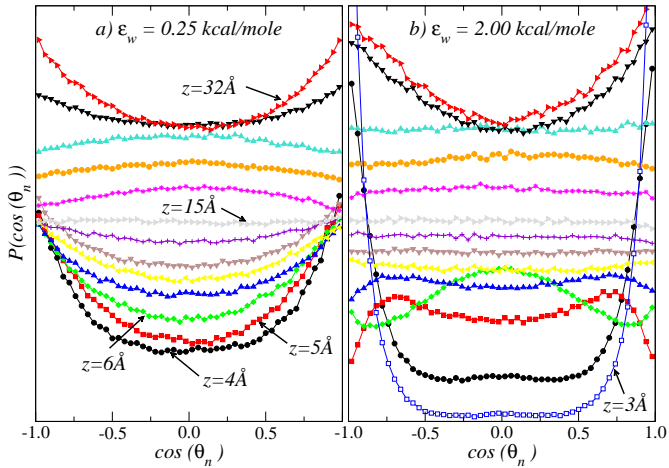}
\caption{
$z$-dependence of the probability distribution of the cosine of the angle $\theta_{n,i}$ between the normalized vector normal to the molecular plane of the water molecules and the vector normal to the support. We have considered the two extreme values $\epsilon_w=0.25, 2.0$ kcal/mole ((a) and b) respectively). A few values for the distance from the support are indicated by arrows and the curves have been shifted vertically for clarity. Same symbols and colors are used at analogous distances in the two cases. A detailed discussion of these data is included in the text.
}
\label{fig:angles}
\end{figure}

We have also investigated in details the presence of orientational order for the water molecules, at different distances from the support. We have subdivided the film in non-overlapping slices with a thickness $\delta z=1$ \AA\ , and have calculated for each slice the probability distribution of $\cos(\theta_{n,i})=\mathbf{u}_i\cdot\mathbf{\hat{z}}$. Here $\theta_{n,i}$  is the angle between the normalized vector normal to the molecular plane of molecule $i$, $\mathbf{u}_i$, and the outward vector normal to the support, $\mathbf{\hat{z}}$. The results for the two limiting cases $\epsilon_w$=0.25 and 2.00 kcal/mole are shown in Fig.~\ref{fig:angles} a) and b), respectively. Data have been shifted vertically for clarity. At short distances (curves on the bottom) we have observed a high degree of orientational order in the hydrophobic case, a), with a high probability associated to water molecules with the molecular plane parallel to the support ($\cos(\theta_{n})=\pm 1$). Note that in these regions the presence of the ionomer is non-negligible. In the hydrophilic case, b), where the ionomer is almost absent, a remarkable modulation is present, with clear minima at $\cos(\theta_{n})\simeq\pm 0.5$ for $z\le 4$\AA\ , followed by an alternation of maxima and minima at $\cos(\theta_{n})\simeq\pm 0.75$, up to $z=8$\AA . Similar results were reported in ~\cite{spohr1989computer}, where the Author studied the rotational order of {\em pure} water in contact with a nano-structured $(100)$ platinum (hydrophilic) surface. This comparison confirms the capability of our mean-field model for the support to grasp features present in much more detailed descriptions. Beyond the interfacial region, the probability of planar configuration is still enhanced in the hydrophobic case, up to the central region of the film ($\simeq 15$\AA), while the hydrophilic case appears more isotropic. Closer to the ionomer/air interfaces, the probability of molecular orientations parallel to the support is again augmented, for both values of $\epsilon_w$. Beyond the differences evident in the two cases, we can conclude for an overall enhanced degree of rotational order in the vicinities of the interfaces for the more hydrophilic supports.  

{\bf Tranport properties.} From the above observations we can predict an impact of changes in morphology on local mass transport properties of water molecules in different regions within the film. These are, indeed, expected to depend on the interplay between confinement due to the hydrophobic matrix and support, and direct interaction with the sulfonate groups. To quantify space-dependent transport, we restrict to the $xy$-plane, slice the film in overlapping slabs of width $\delta z=4$\AA\ equally spaced by $1$\AA\, and consider the following generalized form of the mean-squared displacement:
\begin{equation}
\langle r^2(t,z)\rangle=\frac{1}{N}\sum_{i=1}^N\langle |\mathbf{r}_i(t)-\mathbf{r}_i(0)|^2\delta(z_i(0)-z)\rangle.
\label{eq:msd}
\end{equation}
Here $\langle\rangle$ is the thermodynamic average and $\mathbf{r}_i(t)$ is the $xy$ projection of the three-dimensional position vector of the oxygen atom of water molecule $i=1,\ldots , N$. Only molecules satisfying $z_i(0)=z$ give non-zero contribution to the summation, and  Eq.~(\ref{eq:msd}) corresponds to the dynamics of molecules which, at time $t=0$, were at a distance $z$ away from the support. Note that, at subsequent times, water molecules will change their distance from the support, therefore in general, $z_i(t)\neq z_i(0)$ for $t> 0$.
We are now in the position to define a $z-$dependent in-plane diffusion coefficient, $D(z)$, {\em via} the Einstein relation,
\begin{equation}
D(z)=\lim_{t\rightarrow\infty}\frac{1}{4 t}\langle r^2(t,z)\rangle.
\label{eq:einstein}
\end{equation}
This strategy has been demonstrated to properly characterize space-dependent translational diffusion in confined geometries~\cite{scheidler2002,scheidler2004}.
\begin{figure}[b]
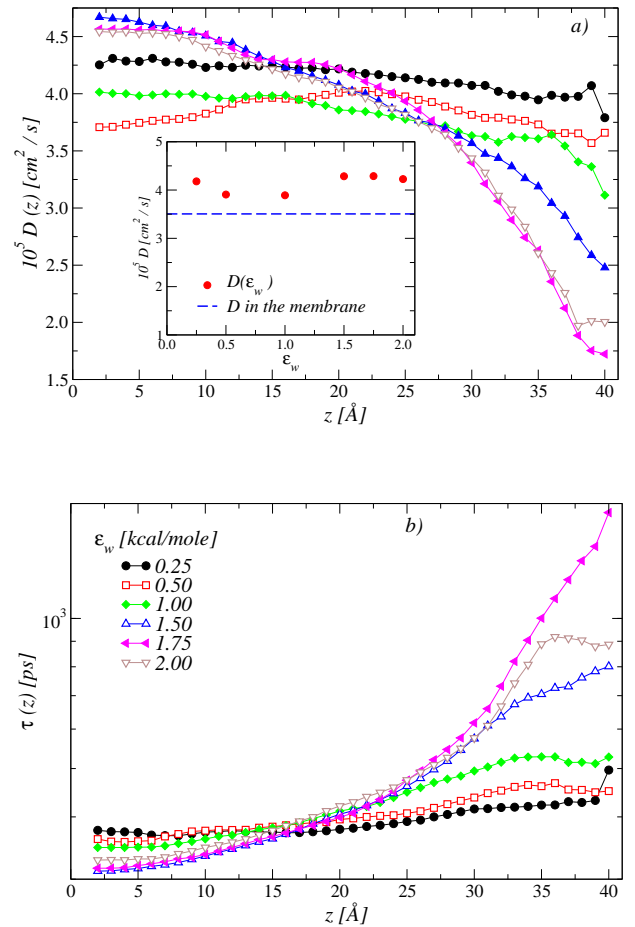

\centering
\includegraphics[width=0.45\textwidth]{diffusion-vs-z.eps}
\vspace{1.0cm}

\includegraphics[width=0.45\textwidth]{tau-vs-z.eps}
\caption{
a) Water molecules diffusion coefficient at the indicated values of $\epsilon_w$. Inset: Total diffusion coefficient, $D$, (circles) integrated on the entire film. $D$ is almost constant on the investigated range, but always higher than the value calculated in the membrane (dashed line). Main panel: $D(z)$, as calculated from Eq.~(\ref{eq:msd}). b) $z-$dependent structural relaxation time for water molecules, extracted from the self-intermediate scattering function.}
\label{fig:diffusion}
\end{figure}

We first show in the inset of Fig.~\ref{fig:diffusion}(a) the total diffusion coefficient for water molecules, integrated in the entire film, compared to the analogous quantity calculated in the membrane, far from any boundaries. (The diffusion coefficient of our pure water model in the same conditions is $\simeq 7.9\times 10^{-5}$~cm$^2$/s.) Interestingly, we observe a diffusion in the film augmented (of about $15\%$) compared to the membrane, at all values of $\epsilon_w$. Note that, although some kind of modulation with $\epsilon_w$ is observed, we cannot conclude any continuous dependence of diffusion on the degree of hydrophilicity of the support. Next, we plot our data as a function of $z$ in the main panel of Fig.~\ref{fig:diffusion}(a). For $\epsilon_w\le 1$~kcal/mole, $D(z) $ is quite uniform across the film, and everywhere close to the average total value. This is consistent with a picture where, at high hydration, water is embedded in a crowded confining environment which is anyhow quite homogeneous within the film. In contrast, for $\epsilon_w>1$~kcal/mole dramatic changes are visible, following the accumulation of the ionomer at intermediate and large distances, with the consequent formation of pure water layers at small $z$. Diffusion becomes strongly heterogeneous, steadily decreasing across the film with a rate which increases with $\epsilon_w$. In the region close to the support, diffusion is enhanced compared to the integrated value, as also reported in the case of simple liquids confined by smooth boundaries~\cite{scheidler2002}. At intermediate distances, $D(z)$ is close to the integrated values, becoming strongly suppressed at higher distances ($z>25$\AA), where interactions with polymer backbones and sulfonate groups are very strong.
\begin{figure}[t]
\centering
\includegraphics[width=0.49\textwidth]{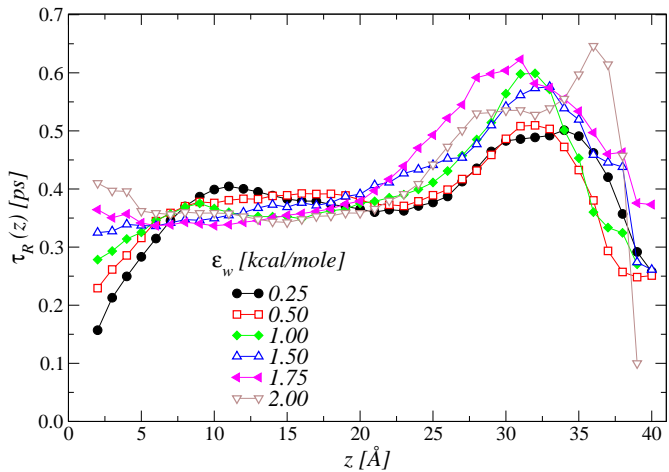}
\caption{
$z-$dependent rotational relaxation time, extracted from an appropriate time-correlation function of the unit vectors normal to the water molecular plane. It is 
clear a strong correlation with the sulfonate groups mass distribution of Fig.~\ref{fig:masses}(c).
}
\label{fig:legendre}
\end{figure}

We have checked the validity of these results by extracting a space-dependent structural relaxation time, $\tau(z)$, from a generalized one-particle intermediate scattering function,
\begin{equation}
F_q(t,z)=\frac{1}{N}\sum_{i}\langle\exp{i\mathbf{q}\cdot[\mathbf{r}_i(t)-\mathbf{r}_i(0)]}\delta(z_i(0)-z)\rangle,
\label{eq:fqt}
\end{equation}
where $\mathbf{q}$ is the wave-vector. We have considered the smallest 2-dimensional $\mathbf{q}$ compatible with the simulation box, $|\mathbf{q}|=2\pi/L_x\simeq 0.08$\AA$^{-1}$. The relaxation time is calculated from the often used relation $F_q(\tau(z),z)=1/e$.
Our findings are shown in Fig.~\ref{fig:diffusion}(b) and support our interpretation of diffusion data. We note that the apparent inversion of the two curves at largest $\epsilon_w$'s can be ascribed to difficult sampling at large $z$, with no qualitative impact on our conclusions.

Even more interesting is the effect of film nano-structuration on orientational dynamics. In Fig.~\ref{fig:legendre} we show the rotational relaxation time, $\tau_R(z)$, extracted from the correlation function,
\begin{equation}
C_2(t,z)=1/N\sum_i\langle P_2(\mathbf{u}_i(t)\cdot\mathbf{u}_i(0))\delta(z_i(0)-z)\rangle.
\label{eq:legendre}
\end{equation}
 $\mathbf{u}_i$ is, as above, the vector normal to the molecular plane of water molecule $i$, and $P_2(x)$  are the second order Legendre polynomials. This correlation function is customary used to characterize the orientational dynamics of molecular liquids and can be measured in light scattering experiments. Here the effect of the interaction of water molecules dipoles with sulfonate groups is striking, with a non-monotonic modulation of the rotational relaxation time which strictly follows the $SO_3^-$ mass distributions of Fig.~\ref{fig:masses}(c). We have checked that other cases (in particular, $l=1$) give very similar qualitative results, showing different average relaxation times but consistent space-dependent behaviour.

\section{Conclusions} 
We have studied by classical Molecular Dynamics simulation the formation of Nafion thin-films at the interface with unstructured supports, characterized by their global wetting properties only. By tuning a single control parameter, the strength of the interaction with the support,  $\epsilon_w$, we have been able to investigate in a unique framework a variety of environments peculiar of the catalyst layer, ranging from hydrophobic to hydrophilic (representative of carbon and platinum, respectively). 

We have found that an increase of the hydrophilic character of the support has strong impact on the conformation of the film, which transforms from an irregular lamellar phase (where an extended water pool is sandwiched by ionomer sheets) to a phase-separated configuration, where water floods the interface with the support and polymers accumulate at the top. Also, we have observed a well developed orientational order for water molecules at short distances from both the ionomer/support and ionomer/air interfaces, especially in the more hydrophilic cases. This order is suppressed in the middle of the film, as expected. Intriguingly, these mutations do not have strong impact on {\em average} transport features of water molecules, as measured from the diffusion coefficient integrated on the entire film. In contrast, by implying a more involved analysis of diffusion at different distances from the support, we have discovered that dynamics at large $\epsilon_w$ is highly heterogeneous across the film. We have shown that the diffusion coefficient is enhanced above the average value in regions close to the support, and strongly suppressed close to the ionomer/air interface, with a rate continuously controlled by $\epsilon_w$. Moreover, very close correlation exists between rotational dynamics and spatial sulfonate groups distribution. 

We expect a significant impact of our work on the interpretation of experimental results. Indeed, in our Molecular Dynamics simulations we have access to all available degrees of freedom of the system, at the atomic level. This is in contrast with large part of modern experimental techniques, which measure properties averaged on quite large real-space domains, often providing wave-vectors dependent spectra. Our data can therefore be, in some cases, directly compared with experiments or, most importantly, to be used to rationalize and clarify exceedingly coarse-grained information. Note that, in the case of Nafion thin films, this possibility is even more beneficial, due to a corpus of available experimental data  which is still quite limited. Experimental techniques which have been employed for the characterization of Nafion thin films include Transmission Electron Microscopy (TEM)~\cite{modestino2013self}. This is able to provide a quite coarse-grained picture of the existing ionic domains, possibly qualitatively comparable with the snapshots of Fig.~\ref{fig:snapshots}. Due to the well extended wave-vectors range available in our simulation boxes, a more quantitative comparisons of our data is possible with detailed information about the topology of the ionic domains coming from Neutron reflectivity experiments~\cite{eastman2012}, or Grazing-Incidence Small-Angle X-ray Scattering (GISAXS)~\cite{bass2011,eastman2012,modestino2013self}. In this case, the mass distributions of the different chemical species shown in Fig.~\ref{fig:masses} can be used for devising the structural fitting models needed for the interpretation of experimental data. Also, modern Nuclear Magnetic Resonance techniques~\cite{li2011linear}, provide information very similar to the rotational observable considered in Eq.(\ref{eq:legendre}). 

This work could also have implications on our understanding of the interplay transport/nano-structure in Nafion {\em membranes (bulk)}. Details of proton and water transport in the ionic domains confined by the Nafion ionomer matrix are still an open issue~\cite{kreuer2004transport}. Recent Quasielastic Neutron Scattering (QENS) spectra~\cite{perrin2007quasielastic} have been rationalized in terms of populations of water/hydronium  molecules with different (fast/slow) dynamics. It is tempting to speculate about the possibility that those populations could be the manifestation of the inhomogeneous dynamical properties described in this work. Indeed, we are convinced that configurations as the bottom snapshot of Fig.~\ref{fig:snapshots} could be considered as (regularized) models for the {\em local} morphology of ionic domains in the membrane, with the support constituting the symmetry plane of the channel available for transport. Work is in progress to explore this conjecture by directly probing inhomogeneous transport in the membrane. 

Finally, we believe that our observations provide useful information to be taken into account in the optimization of actual fuel cell catalyst layers. 

\appendix*
\section{Methods}

{\bf Simulation details.} We have simulated by classical Molecular Dynamics simulation systems formed by $20$ polymers chains, containing $10$ pendant side-chains each. Electro-neutrality was kept by adding $200$ hydronium molecules. We have considered high hydration condition $\lambda=22$, where $\lambda$ is the number of water molecules per sulfonic acid group. Periodic boundary conditions were imposed in the $xy$-plane containing the support, while open boundaries were kept in the $z$-direction. We have fixed a temperature $T=350$~K, and determined the box size ($L_x=L_y\simeq 80$ \AA) by matching a pressure $P\simeq0$~atm. Standard Berendsen thermostat and pressostat have been used. 

Less conventional computing details include: calculation of long-range Coulomb interactions in slab geometry; thermostat coupled to the $xy$-components of velocities only; detection and deletion in subsequent data analysis of molecules evaporated during the molecular dynamics trajectory (a molecule is considered as evaporated, and therefore disregarded, if at any time the $z$-coordinate exceeds $z_{max}=$~50 \AA\ , which is the typical thin film thickness). For production runs we have used the large-scale parallel Molecular Dynamics simulation tool LAMMPS~\cite{plimpton1995fast}.

{\bf System preparation.}
The impact of the fabrication process on morphology and properties of Nafion thin and ultra-thin films, both free-standing and supported, is a very investigated problem (see, among others, \cite{paul2013characteristics}). Although these issues are not considered in this work, we were confronted to the choice of the initial preparation procedure to follow. We decided to initialize the system by randomly placing in the entire available simulation box volume both solvent molecules (water and hydronium) and the ionomer. We followed the system evolution at high temperature and ambient pressure, for a time sufficient to solve all initial overlaps and loose any memory of the initial configuration. Next, we gently annealed the system to ambient temperature, letting it to self-organize and deposit on the substrate. The production runs were started when a stable morphology was reached, as checked by controlling stationarity of the main observables. The same preparation procedure was performed for all systems investigated. We are convinced that this protocol minimizes any bias on morphology formation, which is therefore exclusively controlled by the interplay of the different interaction forces. Compared to our present results, we only expect {\em quantitative} discrepancies in the case of any other preparation protocol where the different chemical species are mixed with no evident initial order or symmetry. We obviously anticipate {\em qualitative} differences in the case of any other biased preparation, {\em e.g.}, adding separately and/or at different times the ionomer and solvent molecules.

{\bf Contact angle calculations.} For selected values of $\epsilon_w\in[0.25,2]$~kcal/mole we have calculated the values of the contact angle of a droplet of water molecules interacting with the support {\em via} Eq.~(\ref{equation:wall}), and demonstrated that this range encompasses hydrophobic $(\theta_c>\pi/2)$ to strongly hydrophilic $(\theta_c<\pi/2)$ behaviour. In order to determine the values of the contact angle corresponding to different $\epsilon_w$'s we have performed an additional series of simulations of spherical clusters formed by $3500$ SPC/E water molecules in vacuum conditions, in the $NVT$ ensemble. After formation the cluster was put in contact with the support and the contact angles have been estimated by directly fitting the contour shape of the droplets~\cite{shi2009molecular,werder2003water}. We have therefore considered a cylindrical region, normal to the support and encompassing the entire droplet, with the principal axis passing through its center of mass. Next we have considered bins in the $z-$direction with a width of $5$\AA\ and separated by $2$\AA\ . For each bin we have calculated the horizontal mass density profile, {\em i.e.}, the water density in the slab as a function of the local drop radius $R_{drop}$ parallel to the support.  $R_{drop}$ is the distance from the center of the cylinder at which the density falls below $0.2 \, g/cm^3$. These profiles are shown in Fig.~\ref{fig:contact}a). To obtain the water contact angle from these data we assume that the curves form a segment of circumference and perform a circular fit to the data. An extrapolation to the support surface provides us with the values of $\theta_c$~\cite{shi2009molecular,werder2003water}.
\bibliography{references}
\end{document}